\documentstyle[12pt]{article}
\newcommand{\ltwid}{\raise.3ex\hbox{$<$\kern-.75em\lower1ex\hbox{$\sim
$}}}

\title{{\hfill\normalsize ITP-94-58E}\\[1.0cm]
Dimensional Reduction and Dynamical Chiral Symmetry Breaking
by a Magnetic Field in $3+1$ Dimensions}

\author{{\sl V.P. Gusynin , V.A. Miransky  and I.A. Shovkovy}\\
{\sl Bogolyubov Institute for Theoretical Physics,
252143 Kiev, Ukraine}}
\date{}
\begin{document}
\maketitle

\vfill

\begin{abstract}
It is shown that in $3+1$ dimensions, a constant magnetic field is a
catalyst of dynamical chiral symmetry breaking, leading to generating a
fermion mass even at the weakest attractive interaction between fermions.
The essence of this effect is the dimensional reduction $D \rightarrow D-2$
($3+1 \rightarrow 1+1$) in the dynamics of fermion pairing in a magnetic field.
The effect is illustrated in the Nambu-Jona-Lasinio model.
Possible applications of this effect are briefly discussed.
\end{abstract}

\vfill
\eject

Recently it has been shown that a constant magnetic field acts as a
strong catalyst of dynamical symmetry breaking  in $2+1$ dimensions,
leading to generating a fermion mass even at the
weakest attractive interaction between fermions \cite{GMS}. In this paper
we will show that a similar effect takes place in $3+1$ dimensions. The
essence of this effect is the dimensional reduction $D \rightarrow D-2$
of the dynamics of fermion pairing in a magnetic field: while at $D=2+1$
the reduction is $2+1 \rightarrow 0+1$, at $D=3+1$ it is $3+1 \rightarrow
1+1$. As we shall see, this leads to dynamical chiral symmetry breaking
even at the weakest attractive interaction between fermions.
We stress that this effect is universal, {\em i.e.,\/}
model independent.

We begin by considering the basic points in the problem of a relativistic
fermion in
a constant magnetic field $B$ (directed along the $x_3$ coordinate)
in $3+1$ dimensions. The Lagrangian density
is
\begin{equation}
{\cal L} = \frac{1}{2} \left[\bar{\Psi},(i\gamma^\mu D_\mu-m)\Psi\right],
\end{equation}
where the covariant derivative $D_\mu$ is
\begin{equation}
D_\mu=\partial_\mu-ie A^{ext}_\mu,\qquad A^{ext}_\mu =-B x_2 \delta_{\mu1}.
\end{equation}
The energy spectrum of fermions is \cite{Akh}:
\begin{equation}
E_n(k_3) = \pm\sqrt{m^2+2|eB|n+k^2_3},\ \  n=0,1,2,\dots .
\end{equation}
(the Landau levels). Each Landau level is degenerate: at each value
of the momentum $k_3$, the number of states equals
$\frac{|eB|}{2\pi}S_{12}$ and $\frac{|eB|}{\pi}S_{12}$ at $n=0$ and $n>0$,
respectively ($S_{12}$ is the square in the $x_1 x_2$-plane). The
degeneracy is connected with the momentum $k_1$ which at the same time
coincides with the $x_2$-coordinate of the center of a fermion orbit in the
$x_1 x_2$-plane \cite{Akh}.

As the fermion mass $m$ goes to zero, there is no energy gap between the
vacuum and the lowest Landau level (LLL) with $n=0$. The density of the
number of states of fermions on the energy surface with $E_0=0$ is
\begin{equation}
\nu_0 =\left. V^{-1}\frac{dN_0}{dE_0} \right|_{E_0=0}=S^{-1}_{12}L^{-1}_3
\left. \frac{dN_0}{dE_0}\right|_{E_0=0}=\frac{|eB|}{4\pi^2}
\end{equation}
where $E_0=|k_3|$ and $dN_0=S_{12}L_3\frac{|eB|}{2\pi}
\frac{dk_3}{2\pi}$ (here $L_3$ is the size in $x_3$-direction). We will see
that the dynamics of the LLL plays the crucial role in catalyzing
spontaneous chiral symmetry breaking. In particular, the density $\nu_0$
plays the same role here as the density of states on the Fermi surface
$\nu_F$ in the theory of superconductivity \cite{BCS}.

The important point is that the dynamics of the LLL is essentially $(1+1)$-
dimensional. Indeed, it is described by two continuous variables: the
momentum $k_3$ and the energy E. This is because, in a magnetic field, the
momentum $k_2$ of free fermions is replaced by the discrete quantum number
$n$ and, besides that, the dynamics does not depend on the momentum $k_1$
(the degeneracy in $k_1$). Thus there is the dimensional reduction
$D \rightarrow D-2$ in the LLL dynamics.

In order to see this dimensional reduction explicitly, let us calculate
the chiral condensate $\langle 0|{\bar\Psi}\Psi|0 \rangle$ in this problem.
The condensate is expressed through the fermion propagator
$S(x,y)=\langle 0|T\Psi(x){\bar\Psi}(y)|0 \rangle$:
\begin{equation}
\langle0|\bar{\Psi}\Psi|0\rangle=-\lim_{x\to y} trS(x,y).
\end{equation}
The propagator of a fermion in a magnetic field was calculated by Schwinger
long ago \cite{Sch}:
\begin{equation}
S(x,y) = \exp (ie\int^x_y A_\lambda^{ext} dz^\lambda) \tilde{S} (x-y),
\end{equation}
\begin{eqnarray}
\tilde{S} (x) &=& -i\int^\infty_0 \frac{ds}{16(\pi s)^{2}}
e^{-ism^2} e^{-\frac{i}{4s}\left((x^0)^2-{\bf x}^2_{\perp}(eBs)cot(eBs)-
(x^3)^2\right)}.\nonumber\\
&\cdot& \Bigg(m+\frac{1}{2s}\left[ \gamma^0x^0-(\gamma^1x^1+
\gamma^2x^2)eBs\cot(eBs)-\gamma^3x^3 \right] -
\frac{eB}{2}\left[ \gamma^2x^1-\gamma^1x^2 \right] \Bigg)
\nonumber \\
&\cdot& \Bigg(esB\cot(eBs) - \gamma^1 \gamma^2 (eBs)\Bigg),
\end{eqnarray}
where ${\bf x}^2_{\perp}=(x^1,x^2)$. The integral in Eq.~(6) is calculated
along the straight line.

The Fourier transform of $\tilde{S}$ in Euclidean space (with
$k^0\to ik_4, s\to -is$) is
\begin{eqnarray}
\tilde{S}_E(k) &=&-i \int^\infty_0 ds \exp \left[-s\Bigg(m^2+k^2_4+k^2_3
+{\bf k}^2_{\perp}
\frac{\tanh(eBs)}{eBs}\Bigg)\right].
\nonumber \\
&\cdot& \left(-k_\mu\gamma_\mu+m+\frac{1}{i}(k_2\gamma_1-k_1\gamma_2)
\tanh(eBs)\right) \left(1+\frac{1}{i}\gamma_1\gamma_2\tanh(eBs)\right)
\end{eqnarray}
(here ${\bf k}_{\perp}=(k_1,k_2)$, $\gamma_4=-i\gamma^0$,
$\gamma_i\equiv\gamma^i$ $(i=1,2,3)$ are
antihermitian matrices). From Eqs.(5),(6) and (8) we find the following
expression for the condensate:
\begin{eqnarray}
\langle0|\bar{\Psi}\Psi|0\rangle & = & - \frac{i}{(2\pi)^4} tr \int d^4 k
\tilde{S}_E (k)
\nonumber\\
&=& - \frac{4m}{(2\pi)^4}\int d^4 k
\int^\infty_{1/\Lambda^2}ds
\exp\left[-s\left(m^2+k_4^2+k_3^2+{\bf k}^2_{\perp} \frac{\tanh(eBs)}{eBs}
\right) \right]
\nonumber\\
&=& -eB\frac{m}{4\pi^{2}} \int^\infty_{1/\Lambda^2} \frac{ds}{s}
e^{-sm^2} \coth(eBs)
\stackrel{m\to0}{\rightarrow} -|eB| \frac{m}{4\pi^2}\left(
\ln\frac{\Lambda^2}{m^2}+O(m^0)\right),
\end{eqnarray}
where $\Lambda$ is an ultraviolet cutoff. One can see that a singular,
$\ln(1/m^2)$, behavior of the integral in Eq.(9) is formed at large,
$s \rightarrow \infty$, distances ($s$ is the proper time coordinate).
Actually one can see from Eq.(9) that the magnetic field removes the two space
dimensions in the infrared region thus reducing the dynamics to a $(1+1)$-
dimensional dynamics which has more severe infrared singularities and leads
to the logarithmic singularity in the condensate.

Let us show that this singularity appears due to the LLL dynamics.
For this purpose, by using the identity $tanh(x)=1-2exp(-2x)/[1+exp(-2x)]$
and the relation \cite{Ryz}:
\begin{equation}
(1-z)^{-(\alpha+1)}\exp\left(\frac{xz}{z-1}\right)=\sum_{n=0}^{\infty}
L_n^{\alpha}(x)z^n,
\end{equation}
where $L_n^{\alpha}(x)$ are the generalized Laguerre polynomials, the
propagator $\tilde{S}_E (k)$ can be decomposed over the Landau level
poles \cite{Cho}:
\begin{equation}
\tilde{S}_E (k)=-iexp\left(-\frac{{\bf k}^2_{\perp}}{eB}\right)
\sum_{n=0}^{\infty}(-1)^n\frac{D_n(eB,k)}{k_4^2+k_3^2+m^2+2eBn}
\end{equation}
with
\begin{eqnarray*}
D_n(eB,k)&=&(m-k_4\gamma_4-k_3\gamma_3)\left[(1-i\gamma_1\gamma_2)
L_n(2\frac{{\bf k}^2_{\perp}}{eB})-(1+i\gamma_1\gamma_2)L_{n-1}
(2\frac{{\bf k}^2_{\perp}}{eB})\right]\\
&+&4(k_1\gamma_1+k_2\gamma_2)L_{n-1}^1(2\frac{{\bf k}^2_{\perp}}{eB}),
\end{eqnarray*}
where $L_n \equiv L_n^0$ and $L_{-1}^{\alpha}(x)=0$ by definition.
This relation implies that the logarithmic
singularity in the condensate appeares due to the LLL. It also explicitly
demonstrates the $(1+1)$-dimensional character of the LLL dynamics.

The above consideration suggests that there is a universal mechanism of the
enhancement of generating fermion masses in a magnetic field in $3+1$
dimensions: the fermion pairing takes place essentially for fermions at the
LLL and this pairing dynamics is $(1+1)$-dimensional (and therefore strong)
in the infrared region.
This in turn suggests that in a magnetic field spontaneous chiral symmetry
breaking takes place even at the weakest attractive interaction between
fermions in $3+1$ dimensions. Actually we shall see that the energy surface
$E_0=0$ in the LLL in chiral invariant theories plays the same role as the
Fermi surface in BCS theory of superconductivity (and the density $\nu_0$ (4)
plays the role of the density of states on the Fermi surface).

Let us consider the Nambu-Jona-Lasinio (NJL) model with the $U_L(1)\times
U_R(1)$ chiral symmetry:
\begin{equation}
{\cal L} = \frac{1}{2} \left[\bar{\Psi}, (i\gamma^\mu D_\mu)\Psi\right] +
\frac{G}{2} \left[ (\bar{\Psi}\Psi)^2+(\bar{\Psi}i\gamma^5\Psi)^2 \right],
\end{equation}
where $D_\mu$ is the covariant derivative  (2)  and fermion fields carry an
additional, ``color", index $\alpha=1,2,\dots,N$. The theory is equivalent
to the theory with the Lagrangian density
\begin{equation}
{\cal L}=\frac{1}{2} \left[\bar{\Psi}, \left( i\gamma^\mu D_\mu\right)
 \Psi\right] - \bar{\Psi}(\sigma+i\gamma^5\pi)\Psi
- \frac{1}{2G} \left(\sigma^2+\pi^2\right)
\end{equation}
$ \left(\sigma=- G\bar{\Psi}\Psi, \pi=-G\bar{\Psi}i\gamma^5\Psi\right )$.
The effective action for the composite fields $\sigma$ and $\pi$ is:
\begin{equation}
\Gamma(\sigma,\pi) = -\frac{1}{2G}\int d^4x(\sigma^2+\pi^2) +
\tilde{\Gamma}(\sigma,\pi),
\end{equation}
where
\begin{equation}
\tilde{\Gamma}(\sigma,\pi) =- i Tr Ln \left[i\gamma^\mu D_\mu
- (\sigma+i\gamma^5\pi)\right].
\end{equation}
As $N\to\infty$, the path integral over the composite fields is dominated
by stationary points of their action:
$\delta\Gamma/\delta\sigma=\delta\Gamma/\delta\pi=0$. We
will analyze the dynamics in this limit by using the expansion of the action
$\Gamma$ in powers of derivatives of the composite fields.

We begin by calculating the effective potential $V$. Since $V$ depends
only on the $U_L(1)\times U_R(1)$-invariant $\rho^2=\sigma^2+\pi^2$, it is
sufficient to
consider a configuration with $\pi=0$ and $\sigma$ independent of $x$.
Then, using the proper-time method [4], we find the potential from
Eqs.~(6), (7), and (14),(15):
\begin{eqnarray}
V(\rho)&=&\frac{\rho^2}{2G}+\tilde{V}(\rho)=\frac{\rho^2}{2G} +
\frac{N}{8\pi^2} \int^\infty_{1/\Lambda^2} \frac{ds}{s^2}
e^{-s\rho^2} \frac{1}{l^2}\coth(\frac{s}{l^2})
\nonumber\\
&=&\frac{\rho^2}{2G}+\frac{N}{8\pi^2} \Bigg[\frac{\Lambda^4}{2}+
\frac{1}{3l^4}\ln(\Lambda l)^2 + \frac{1-\gamma-\ln2}{3l^4}-(\rho\Lambda)^2
\nonumber\\
&+&\frac{\rho^4}{2}\ln(\Lambda l)^2+\frac{\rho^4}{2}(1-\gamma-\ln2)+
\frac{\rho^2}{l^2}\ln\frac{\rho^2l^2}{2}-\frac{4}{l^4}\zeta'(-1,
\frac{\rho^2l^2}{2}+1)\Bigg]
\end{eqnarray}
where the magnetic length $l\equiv|eB|^{-1/2}$, $\zeta'(-1,x)=
\frac{d}{d\nu}\zeta(\nu,x)|_{\nu=-1}$, $\zeta(\nu,x)$ is the generalized
Riemann zeta function \cite{Ryz}, and $\gamma\approx0.577$ is the Euler
constant. The gap equation $dV/d\rho=0$ is:
\begin{equation}
\Lambda^2\left(\frac{1}{g}-1\right)=-\rho^2\ln\frac{(\Lambda l)^2}{2}+
\gamma\rho^2+l^{-2}\ln\frac{(\rho l)^2}{4\pi}+2l^{-2}\ln\Gamma\left(
\frac{\rho^2 l^2}{2}\right)+O\left(\frac{1}{\Lambda}\right),
\end{equation}
where the dimensionless coupling constant $g=NG\Lambda^2/(4\pi^2)$. In the
derivation of this equation, we used the relations \cite{Ryz}
\begin{equation}
\frac{d}{dx}\zeta(\nu,x)=-\nu\zeta(\nu+1,x),
\end{equation}
\begin{equation}
\frac{d}{d\nu}\zeta(\nu,x)|_{\nu=0}=\ln\Gamma(x)-\frac{1}{2}\ln2\pi,
\qquad \zeta(0,x)=\frac{1}{2}-x.
\end{equation}
As $B\to0$ ($l\to\infty$), we recover the known gap equation in the NJL
model (for a review see the book \cite{Mir}):
\begin{equation}
\Lambda^2\left(\frac{1}{g}-1\right)=-\rho^2\ln\frac{(\Lambda)^2}{\rho^2}.
\end{equation}
This equation admits a nontrivial solution only if $g$ is supercritical,
$g>g_c=1$ (as Eq.(13) implies, a solution to Eq.(20), $\rho=\bar{\sigma}$,
coincides with the fermion dynamical mass, $\bar{\sigma}=m_{dyn}$, and the
dispersion relation for fermions is Eq.(3) with $m$ replaced by
$\bar{\sigma}$). We will show that the magnetic field changes the situation
dramaticaly: at $B\neq0$, a nontrivial solution exists at all $g>0$.

We shall first consider the case of subcritical $g$, $g<g_c=1$, which
in turn can be divided into two subcases: {\cal a})$g\ll g_c$ and
{\cal b})$g\rightarrow g_c-0$ (nearcritical $g$). Since at $g<g_c=1$
the left-hand
side in Eq.(17) is positive and the first term on the right-hand side in
this equation is negative, we conclude that a nontrivial solution to this
equation may exist only at $\rho^2 \ln\Lambda^2 l^2\ll l^{-2}
\ln(\rho l)^{-2}$. Then at $g\ll 1$, we find the solution:
\begin{equation}
m^2_{dyn}\equiv\bar{\sigma}^2=\frac{1}{\pi l^2}\exp\left(
-\frac{\Lambda^2l^2}{g}\right)=\frac{eB}{\pi}\exp\left(
-\frac{4\pi^2}{eB N G}\right) .
\end{equation}
It is instructive to compare this result with that in $(2+1)$-dimensional
NJL model in a magnetic field \cite{GMS} and with the BCS relation for the
energy gap in the theory of superconductivity \cite{BCS}. While the expression
(21) for $m_{dyn}^2$ has an essential singularity at $g=0$, in $(2+1)$-
dimensional NJL model $m_{dyn}^2$ is analytic at $g=0$: $m_{dyn}^2\sim l^{-4}
\Lambda^{-2}g^2$ \cite{GMS}. The latter is connected with the fact that in
$2+1$
dimensions the condensate $\langle0|\bar{\Psi}\Psi|0\rangle$ is nonzero
even for free fermions in a magnetic field \cite{GMS}. As a result, the
dynamical mass is essentially perturbative in $g$ in that case. This is in
turn connected with the point that because of the dimensional reduction
$D\rightarrow D-2$ in a magnetic field, the dynamics of the fermion pairing
in a magnetic field in $2+1$ dimensions is $(0+1)$-dimensional.

On the other hand, the dynamics of the fermion pairing in a magnetic field
in $3+1$ dimensions is $(1+1)$-dimensional. We recall that, because of the
Fermi surface, the dynamics of the electron pairing in BCS theory is also
$(1+1)$-dimensional. This analogy is rather deep. In particular, the
expression (21) for $m_{dyn}$ can be rewritten in a form similar to that for
the energy gap $\Delta$ in BCS theory: while $\Delta\sim\omega_D
\exp\left(-Const/G_S\nu_F\right)$, where $\omega_D$ is the Debye
frequency, $G_S$ is a coupling constant and $\nu_F$ is the density of states
on the Fermi surface, the mass $m_{dyn}$ is $m_{dyn}\sim \sqrt{eB}
\exp\left(-1/2 N G \nu_0 \right)$ where the density of states $\nu_0$ on the
energy surface $E=0$ of the LLL is given in Eq.(4). Thus the
energy surface $E=0$ plays here the role of the Fermi surface.

Let us now consider the nearcritical $g$ with $\Lambda^2(
1-g)/g\rho^2\ll\ln\Lambda^2l^2$. Looking for a solution with
$\rho^2l^2\ll 1$, we come to the following equation:
\begin{equation}
\frac{1}{\rho^2l^2}\ln\frac{1}{\pi\rho^2l^2}\simeq\ln\Lambda^2l^2,
\end{equation}
{\em i.e.,\/}
\begin{equation}
m_{dyn}^2=\bar{\sigma}^2\simeq|eB|\frac{\ln\left[(\ln\Lambda^2l^2)/\pi
\right]}{\ln\Lambda^2l^2}.
\end{equation}
What is the physical meaning of this relation? Let us recall that at $g=g_c$,
the NJL model is equivalent to the (renormalizable) Yukawa model \cite{Mir}.
In leading order in $1/N$, the renormalized Yukawa coupling $\alpha_Y^{(l)}=
(g_Y^{(l)})^2/(4\pi)$, relating to the energy scale $\mu\sim l^{-1}$, is
$\alpha_Y^{(l)}\simeq\pi/\ln\Lambda^2l^2$ \cite{Mir}. Therefore,the dynamical
mass (23) can be rewritten as
\begin{equation}
m_{dyn}^2\simeq|eB|\frac{\alpha_Y^{(l)}}{\pi}\ln\frac{1}{\alpha_Y^{(l)}}.
\end{equation}
Thus, as have to be in a renormalizable theory, cutoff $\Lambda$ is removed,
through the renormalization of parameters (the coupling constant in this
case), from the observable $m_{dyn}$.

At $g>g_c$, an analytic expression for $m_{dyn}$ can be obtained at weak
magnetic field, satisfying the condition $|eB|^{1/2}/m^{(0)}_{dyn}
\ll1$, where $m^{(0)}_{dyn}$ is the solution to the gap equation (20) with
$B=0$. Then we find from Eq.(17):
\begin{equation}
m_{dyn}^2\simeq\left(m^{(0)}_{dyn}\right)^2\left[1+\frac{|eB|^2}
{3(m^{(0)}_{dyn})^4\ln(\Lambda/m^{(0)}_{dyn})^2}\right],
\end{equation}
{\em i.e.,} $m_{dyn}$ increases with $|B|$\footnote{The fact that in the
supercritical phase of the NJL model, $m_{dyn}$ increases with $|B|$ has
been already pointed out by several authors (for a review see
Ref.~\cite{Klev}).}. In the nearcritical region, with $g-g_c\ll 1$, this
expression for $m_{dyn}^2$ can be rewritten as
\begin{equation}
m_{dyn}^2\simeq\left(m^{(0)}_{dyn}\right)^2\left[1+\frac{1}{3\pi}
\alpha^{(m_{dyn})}_Y \frac{|eB|^2}
{(m^{(0)}_{dyn})^4}\right],
\end{equation}
where, in leading in $1/N$, $\alpha_Y^{(m_{dyn})}\simeq\pi/\ln(\Lambda/
m_{dyn}^{(0)})^2$ is the renormalized
Yukawa coupling relating to the scale $\mu=m_{dyn}^{(0)}$.

Before turning to calculating the kinetic term in the effective action (that
will allow to describe the Nambu-Goldstone bosons in this problem), let us
consider the question whether the dynamical reduction $3+1\to1+1$ is
consistent with spontaneous chiral symmetry breaking in this problem. This
question is natural since, due to the Mermin-Wagner-Coleman (MWC) theorem
[9,10], there is no spontaneous breakdown of continuous
symmetry at $D=1+1$. However, the MWC theorem is not applicable to our case.
The point is the following. The MWC theorem is based on the fact that gapless
NG bosons cannot exist at $D=1+1$. On the other hand, the reduction $D\to D-2$
in a magnetic field takes place only in charged channels, while the
condensate $\langle0|\bar{\Psi}\Psi|0\rangle$ and the NG bosons are neutral.
Therefore, as we shall see below, unlike the propagator of fermions, the
propagator of the NG bosons has a genuine $(3+1)$-dimensional form, {\em i.
e.} there is no obstacle for a realization of spontaneous chiral symmetry
breaking in a magnetic field in $3+1$ dimensions.

The chiral $U_L(1)\times U_R(1)$ symmetry implies that the general form of
the kinetic term is
\begin{equation}
{\cal L}_k = \frac{F_1^{\mu\nu}}{2} (\partial_\mu\rho_j\partial_\nu \rho_j)
+ \frac{F^{\mu\nu}_2}{\rho^2}(\rho_j\partial_\mu\rho_j)(\rho_i
\partial_\nu\rho_i)
\end{equation}
where $\mbox{\boldmath$\rho$}=(\sigma,\pi)$ and
$F^{\mu\nu}_1$, $F^{\mu\nu}_2$ are
functions of $\rho^2$. We found these functions by using the same method as
in Ref.~\cite{GMS} (see especially Appendix A in the second paper of Ref.~
\cite{GMS}). The result is:
$F^{\mu\nu}_1=g^{\mu\nu}F^{\mu\mu}_1$, $F^{\mu\nu}_2=g^{\mu\nu}F^{\mu\mu}_2$
with
\begin{eqnarray}
F^{00}_1 &=& F^{33}_1=\frac{N}{8\pi^2}\left[\ln\frac{\Lambda^2l^2}{2}-
\psi\left(\frac{\rho^2l^2}{2}+1\right)+\frac{1}{\rho^2l^2}-\gamma+
\frac{1}{3} \right], \nonumber\\
F^{11}_1 &=& F^{22}_1=\frac{N}{8\pi^2}\left[\ln\frac{\Lambda^2}{\rho^2}-
\gamma+\frac{1}{3}\right], \nonumber\\
F^{00}_2 &=&F^{33}_2=-\frac{N}{24\pi^2}\left[\frac{\rho^2 l^2}{2}
\zeta\left(2,\frac{\rho^2 l^2}{2}+1\right)+\frac{1}{\rho^2 l^2}\right], \\
F_2^{11} &=& F^{22}_2=\frac{N}{8\pi^2}\Bigg[\rho^4 l^4
\psi\left(\frac{\rho^2 l^2}{2}+1\right)-2\rho^2 l^2\ln\Gamma\left(
\frac{\rho^2 l^2}{2}\right)\nonumber\\
\qquad &-& \rho^2 l^2 \ln\frac{\rho^2 l^2}{4\pi}-\rho^4 l^4-\rho^2 l^2+1\Bigg]
\nonumber
\end{eqnarray}
where $\psi(x)=d\left(\ln\Gamma(x)\right)/dx$.

We find the following spectrum for the excitations $\sigma$ and $\pi$ from
Eqs.(16),(27) and (28):

{\em a)} subcritical $g$, $g<g_c$, region:

At $g\ll g_c=1$ we find:
\begin{equation}
E_{\sigma}\simeq\left(12m_{dyn}^2+\frac{3(m_{dyn} l)^2(\Lambda l)^2}{g}
{\bf k}^2_{\perp}+k_3^2\right)^{1/2},
\end{equation}
\begin{equation}
E_{\pi}\simeq\left(\frac{(m_{dyn} l)^2(\Lambda l)^2}{g}
{\bf k}^2_{\perp}+k_3^2\right)^{1/2},
\end{equation}
with $m_{dyn}$ defined in Eq.(21). Thus $\pi$ is gapless NG mode. Taking into
account Eq.(21), we find that the transverse velocity $v_{\perp}$ of $\pi$
is less then 1.

In the nearcritical region, we find:
\begin{equation}
E_{\sigma}\simeq\left[4m_{dyn}^2\left(1+\frac{2}{3\ln\pi/
\alpha_Y^{(l)}}\right)+\left(1-\frac{1}{3\ln\pi/\alpha_Y^{(l)}}\right)
{\bf k}^2_{\perp}+k_3^2\right]^{1/2},
\end{equation}
\begin{equation}
E_{\pi}\simeq\left[\left(1-\frac{1}{\ln\pi/
\alpha_Y^{(l)}}\right){\bf k}^2_{\perp}+k_3^2\right]^{1/2},
\end{equation}
(see Eq.(24)). Thus, as should be in a renormalizable theory, the cutoff
$\Lambda$ disappears from observables $E_{\sigma}$ and $E_{\pi}$.

{\em b)} supercritical $g$, $g>g_c=1$, and weak field, $B$, $(l\to\infty)$:
\begin{eqnarray}
E_{\sigma}&\simeq&\Bigg[4m_{dyn}^2\left(1+\frac{1}{3\ln(\Lambda/
m_{dyn})^2}+\frac{4(eB)^2}{9m_{dyn}^4\ln(\Lambda/m_{dyn})^2}\right)
\nonumber\\
&+&\left(1-\frac{11(eB)^2}{45m_{dyn}^4\ln(\Lambda/m_{dyn})^2}\right)
{\bf k}^2_{\perp}+k_3^2\Bigg]^{1/2},
\end{eqnarray}
\begin{equation}
E_{\pi}\simeq\left[\left(1-\frac{(eB)^2}{3m_{dyn}^4\ln(\Lambda/
m_{dyn})^2}\right)
{\bf k}^2_{\perp}+k_3^2\right]^{1/2},
\end{equation}
with $m_{dyn}$ given in Eq.(25). One can see that a magnetic field reduces
the transverse velocity $v_{\perp}$ of $\pi$ in this case as well. Taking
into account that in the nearcritical region, with $g-g_c\ll 1$, the
renormalized Yukawa coupling $\alpha_Y^{(m_{dyn})}$, relating to the
scale $\mu=m_{dyn}$, is
$\alpha_Y^{(m_{dyn})}\simeq\pi/\ln(\Lambda/m_{dyn})^2$ to leading order
in $1/N$, we find in that region:
\begin{equation}
E_{\sigma}\simeq\Bigg[4m_{dyn}^2\left(1+\alpha_Y^{(m_{dyn})}\left(
\frac{1}{3\pi}+\frac{4(eB)^2}{9\pi m_{dyn}^4}\right)\right)
+\left(1-\alpha_Y^{(m_{dyn})}\frac{11(eB)^2}{45\pi m_{dyn}^4}\right)
{\bf k}^2_{\perp}+k_3^2\Bigg]^{1/2},
\end{equation}
\begin{equation}
E_{\pi}\simeq\left[\left(1-\alpha_Y^{(m_{dyn})}\frac{(eB)^2}{3\pi m_{dyn}^4}
\right)
{\bf k}^2_{\perp}+k_3^2\right]^{1/2}.
\end{equation}

The present model illustrates the general phenomenon in $3+1$ dimensions:
in the
infrared region, an external magnetic field reduces the dynamics of fermion
pairing to $(1+1)$-dimensional dynamics (at the lowest Landau level) thus
generating a dynamical mass for fermions even at the weakest attractive
interactions. A concrete sample of dynamical symmetry breaking is of course
different in different models.

In conclusion, let us discuss possible applications of this effect. One
potential application is the interpretation of the results of the GSI
heavy-ion
scattering experiments \cite{Sal} in which narrow peaks are seen in the
energy spectra of emitted $e^{+}e^{-}$ pairs. One proposed explanation
\cite{Celenza} is that a very strong elecromagnetic field, created by
the heavy ions, induces a phase transition in QED to a phase with
spontaneous chiral symmetry breaking. The observed peaks are due to the
decay of positronium-like states in this phase. The effect of the catalysis
of chiral symmetry breaking by a magnetic field, studied in this paper, can
serve as a toy example of such a phenomenon. In order to get a more
realistic model, it would be interesting to extend this analysis to non-
constant background fields. It would be also interesting to consider this
effect in QED. This work is now in progress.

Another, potentially interesting, application can be connected
with the role of iso- and
chromomagnetic backgrounds as models for the QCD vacuum (the Copenhagen
vacuum) \cite{Niel}. Also, as has been suggested recently \cite{DESY},
isomagnetic fields in the vacuum of electroweak Left-Right Models can
induce the parity breakdown. Our work suggests that such field
configurations may play the
important role in triggering chiral symmetry breaking in those theories.

\end{document}